\documentclass{article}
\usepackage{emulateapj}

\usepackage{multirow}
\usepackage{epsf}
\usepackage{rotating}
\usepackage{lscape}
\usepackage{flushrt}
\usepackage{pstricks,apjfonts,amsmath}

\def\asca       {{\em ASCA}\/}
\def\einstein   {{\em Einstein}\/}
\def\rosat      {{\em ROSAT}\/}
\def\g		{{\em g}\/}
\def\r		{{\em r}\/}
\def\i		{{\em i}\/}

\def\B		{{\em B}\/}
\def\deg        {$^{\circ}$}
\def\eg		{{e.g.,}\/}

\begin{document}

\lefthead{MASS TO LIGHT RATIOS OF CLUSTERS}
\righthead{HRADECKY ET AL.}

\title{MASS TO LIGHT RATIOS OF GROUPS AND CLUSTERS OF GALAXIES}

\author{V. Hradecky\altaffilmark{1}, C. Jones\altaffilmark{2},
R.H. Donnelly\altaffilmark{2}, S.G. Djorgovski\altaffilmark{1},
R.R. Gal\altaffilmark{1}, S.C. Odewahn\altaffilmark{3}} 

\altaffiltext{1}{California Institute of Technology, Pasadena, CA
91125} 
\altaffiltext{2}{Harvard-Smithsonian Center for Astrophysics, 60
Garden Street, Cambridge, MA 02138} 
\altaffiltext{3}{Arizona State University, Dept. of Physics \& Astronomy,
Tempe, AZ 85287}

\begin{abstract}

We constrain the mass-to-light ratios, gas mass fractions, baryon mass
fractions and the ratios of total to luminous mass for a sample of eight
nearby relaxed galaxy groups and clusters: A262, A426, A478, A1795,
A2052, A2063, A2199 and MKW4s. We use \asca\ spatially resolved
spectroscopic X-ray observations and \rosat\ PSPC images to constrain
the total and gas masses of these clusters. To measure cluster
luminosities we use galaxy catalogs resulting from the digitization
and automated processing of the second generation Palomar Sky Survey
plates calibrated with CCD images in the Gunn-Thuan \g, \r, and \i\
bands.

Under the assumption of hydrostatic equilibrium and spherical
symmetry, we can measure the total masses of clusters from their
intra-cluster gas temperature and density profiles. Spatially resolved
\asca\ spectra show that the gas temperature decreases with increasing
distance from the center. By comparison, the assumption that the gas
is isothermal results in an underestimate of the total mass at small
radii, and an overestimate at large cluster radii.

We have obtained luminosity functions for all clusters in our
sample. After correcting for background and foreground galaxies, we
estimate the total cluster luminosity using Schechter function fits to
the galaxy catalogs. In the three lowest redshift clusters where we
can sample to fainter absolute magnitudes, we have detected a
flattening of the luminosity function at intermediate magnitudes and a
rise at the faint end. These clusters were fitted with a sum of two
Schechter functions. The remaining clusters were well fitted with a
single Schechter function. 

Assuming $H_0=50~h_{50}~{\rm km}~{\rm s}^{-1}~{\rm Mpc}^{-1}$, the measured
mass-to-light ratios are $\sim 100~h_{50}~{\rm M}_{\odot}/{\rm
L}_{\odot}$. This, along with a high baryonic fraction, is indicative
of a low density universe with $\Omega_0\sim 0.15-0.2$.


\end{abstract}

\section{INTRODUCTION}

Galaxy clusters are the most massive bound systems known and hence
are of interest for investigating cosmological parameters. These can
be constrained by studying fundamental properties such as cluster
mass-to-light ratios, dark matter distributions, gas mass fractions or
the ratios of luminous baryon mass to the total mass. 

The cosmological parameter $\Omega_0$ (the ratio of the mass density
of the universe to the critical density) can be constrained by
measuring the mass-to-light ratios of clusters, estimating the
luminosity density of the universe, and assuming that clusters have a
dark matter content representative of the whole Universe. This
assumption is supported by measurements of the Virgo cluster infall
motion, the cosmic virial theorem (Bahcall et al. 1995) and a weak
gravitational lensing mass estimate of a supercluster of galaxies,
yielding a mass-to-light ratio comparable to that of clusters, ${\rm
M}/{\rm L}=(140\pm20) \, h_{50} ~{\rm M}_{\odot}/{\rm L}_{\odot}$
(Kaiser et al. 1998)

$\Omega_0$ also can be independently constrained by studying the
cluster gas mass fractions (the ratio of the gas mass to total
mass). Predictions from standard big bang nucleosynthesis limit the
baryon density of the universe to $\Omega_b = f_b \Omega_0 =
0.076\pm0.004 h_{50}^{-2}$ (Walker et al. 1991, White et al. 1993,
Tytler et al. 1996, Kirkman et al. 2000), where $f_b$ is the baryon
mass fraction. The luminous baryonic component of clusters consists
primarily of the intra-cluster gas, with a small contribution from
stars. The possible other components not observed result in the
luminous baryons being a lower limit on the baryon fraction. Assuming
that the observed baryon fractions in clusters are representative of
the baryonic content of the whole universe, we can use cluster gas
mass fractions to place an upper limit on $\Omega_0$, given the Hubble
constant $H_0$.

These investigations fundamentally rely on accurate measurements of
cluster luminosities and masses. Summed optical luminosities of
clusters are ideally measured using CCD observations. However,
photometric data sets extending to the cluster virial radii are
available only for relatively small samples for low redshift clusters
(Lopez-Cruz 1995). A presently feasible way of measuring low redshift
cluster luminosities over large volumes is to use photometric CCD
images in a suitable filter system to calibrate photographic survey
plates covering larger areas of the sky. We use the Digitized Second
Palomar Sky Survey photographic plates (Djorgovski et al. 1998),
calibrated with CCD images in the Gunn-Thuan \g, \r, and \i\ bands,
which provide a good match with the plate and filter transmission
curves (Weir et al. 1995a). Palomar Sky Survey plate detection limits
are about $2-3$ magnitudes brighter than for the CCD images. For a
typical low redshift cluster, galaxies with an apparent magnitude
brighter than $\sim 19$ mag contain $>90\%$ of the total cluster
luminosity, making photographic plates well suited for measuring total
cluster light.

Several methods have been used for measuring total cluster masses,
with generally consistent results. This suggests that the total
cluster masses can be measured with reasonable accuracy, although
systematic variations between the different methods do exist.

The oldest method of estimating cluster masses is based on the
distribution of galaxy redshifts (\eg\ the virial mass
estimator). Assuming that the distribution of galaxies is similar to
the distribution of the total mass, the cluster is in virial
equilibrium and the velocity dispersions are isotropic, the virial
mass of a cluster is related to the virial radius, $r_v$, and the line
of sight projected velocity dispersion of galaxies, $\sigma$ by:
$M_v=3\sigma^2r_v/G$. This equation overestimates the total mass if
the cluster is sampled to a radius smaller then $r_v$, since the
surface pressure term in the virial theorem (2U+T=3PV) reduces the
mass needed to bind the system. Moreover, if velocity anisotropies in
the cluster exist, or the assumption that mass follows light does not
hold, the virial mass estimator may produce misleading results (The \&
White 1986, Meritt 1987). For example, Bailey (1982) has shown that
relaxing the mass-follows-light assumption can result in total cluster
masses being considerably reduced. In such a case, a M/L$_V$ ratio as
low as $50\,{h_{50}}$ is consistent with observed velocity dispersions
in the Coma Cluster.

Using X-ray emission from clusters to measure the total masses has
several advantages over virial mass estimators, since some of the
assumptions involved can be observationally tested. Clusters have
X-ray luminosities on the order of $10^{43}-10^{45}$ ergs/sec,
generated primarily by thermal bremsstrahlung from the hot
intra-cluster gas that fills the deep gravitational potential wells
(\eg\ Jones \& Forman 1984). Under the assumptions that this gas is
supported by thermal pressure, is in hydrostatic equilibrium and
spherically symmetric, the total cluster mass can be estimated from
the gas density and temperature profiles (Bahcall \& Sarazin 1977,
Mathews 1978).

Previous optical and X-ray studies of groups and clusters (\eg\ David
et al. 1990, David et al. 1995) generally took into account the
temperature structure for cool systems as measured by
\emph{ROSAT}. For the hottest, richest clusters, the gas was often
assumed to be isothermal and was characterized by the emission
weighted temperature. Crude temperature maps can now be obtained using
the spatially resolved \asca\ spectra, after applying corrections for
the point spread function (PSF) of the \asca\ mirrors. For the
majority of clusters in our sample, we find the temperature declines
with radius. Other studies (Markevitch et al. 1998, Nevalainen et
al. 1999) have also observed declining temperature profiles. Compared
to using the measured temperature profile, the isothermal assumption
underestimates the total mass at small radii, and overestimates it at
large radii. In addition, azimuthal variations in the gas temperatures
have been observed in a number of clusters that are indicative of
recent merger activity (Donnelly et al. 1998, Henriksen et
al. 2000). In such clusters, the assumption of hydrostatic equilibrium
can break down, and the applicability of X-ray mass estimates can be
questioned.

Direct measurements of cluster masses can be obtained from
gravitational lensing distortions of background galaxies. However,
only a limited number of systems have been studied using this method
(\eg\ Smail et al. 1995). For cooling-flow clusters where the
assumption of hydrostatic equilibrium is expected to hold, Allen
(1998) found good agreement between X-ray mass estimates and results
from strong and weak lensing.

In this paper we improve on the earlier measurements of cluster
mass-to-light ratios, gas mass fractions, the limits on the baryon
mass fractions and the constraints on $\Omega_0$ by accounting for the
intracluster gas temperature profiles, as well as using better quality
optical data for measuring cluster luminosities. We study clusters
that show symmetric temperature decline with radius, supporting the
assumption of hydrostatic equilibrium and spherical symmetry. Our
sample consists of 7 clusters (A262, A426, A478, A1795, A2052, A2063,
A2199) and one group (MKW4s). They were selected as members of an
X-ray flux limited sample of clusters that were observed with \asca\
and the \rosat\ PSPC, and are within the limits of the Second Palomar
Sky Survey ($\delta>-3^{\circ}$). The details of our sample are
tabulated in Table~\ref{tab:sample}.


\begin{deluxetable}{ccccccccccc}
\tablecaption{The Sample\label{tab:sample}}
\tablewidth{7in}
\tablehead{Object & RA2000 & DEC2000 & POSS-II field & z\tablenotemark{a} &
Gal. long. & Gal. latt. & E(B-V)\tablenotemark{b}& ${\rm T_x}$(keV)\tablenotemark{c} & $T_{B-M}$\tablenotemark{d}& R\tablenotemark{d}} 
\startdata
A262 & 01:52:50.4 & +36:08:46 & 353, 354  & 0.0161 & 136.59 &
-25.09 & 0.09 & $2.3\pm0.2$ & III & 0 \nl
A426 & 03:18:36.4 & +41:30:54 & 300, 301  & 0.0183 & 150.38 &
-13.38 & 0.18 & $6.2\pm0.4$& II-III & 2 \nl
A478 & 04:13:20.7 & +10:28:35 & 691, 692  & 0.09 & 182.41 &
-28.30 & 0.51 & $8.4^{+0.8}_{-1.4}$ & $\cdots$ & 2 \nl
A1795 & 13:49:00.5 & +26:35:07 & 509, 510 & 0.0616 & 33.79  &
77.16 & 0.01 & $7.8\pm1.0$ & I & 2 \nl
A2052 & 15:16:45.5 & +07:00:01 & 725, 797 & 0.0348 & 9.39   &
50.10 & 0.04 & $2.8\pm0.2$ & I-II & 0 \nl
A2063 & 15:23:01.8 & +08:38:22 & 725      & 0.0354 & 12.85  &
49.71 & 0.03 & $2.3\pm0.2$ & II & 1 \nl
A2199 & 16:28:37.0 & +39:31:28 & 331      & 0.0302 & 62.90  &
43.70 & 0.01 & $4.8\pm0.1$ & I & 2 \nl
MKW4s & 12:06:38.9 & +28:10:18 & 440,441 & 0.0283 & 204.34 &
80.03 & 0.02 & $1.8\pm0.3$ & $\cdots$ & $\cdots$ \nl
\enddata
\tablenotetext{a}{Struble \& Rood (1987), redshift for MKW4s from
Dell'Antonio et al. (1994)} 
\tablenotetext{b}{Schlegel et al. (1998)}
\tablenotetext{c}{Emission-weighted gas temperature with the cooling flow
excluded. Values for A478, A1795 and A2199 are from Markevitch et
al. (1998). Values for remaining clusters are our estimates from
\asca\ analysis.}
\tablenotetext{d}{Bautz-Morgan and Richness classes (Abell et al. 1989).}
\end{deluxetable}

In Section 2 of this paper we discuss the X-ray data reduction and
analysis. In Section 3 we discuss the optical data analysis. The main
results and discussion of their implications are presented in Section
4.  We assume ${\rm H}_0=50 \, h_{50}~{\rm km}~{\rm s}^{-1} {\rm
Mpc}^{-1}$ and ${\rm q}_0=0.5$. All errors are 90\% confidence.

\section{X-ray data reduction and analysis}

Under the assumptions that the intra-cluster medium is spherically
symmetric and in hydrostatic equilibrium supported solely by thermal
pressure, the gas density, $\rho_g$, temperature, $T$, pressure, $p_g$
and mass, $M$, are related by:
\begin{equation}
\frac{d p_g}{d r} = \rho_g\frac{G M(<r)}{r^2}
\end{equation}
\begin{equation}
p_g=\frac{\rho_g k T}{\mu m_p}
\end{equation}
Here $\mu$ is the mean molecular weight of the gas (we assume
$\mu=0.6$), and $k$ is Boltzmann's constant. The mass within a radius
$r$ is then:
\begin{equation}
M(<r)=-\frac{k T(r)}{\mu m_p G}\left(\frac{d\log \rho_g(r)}{d\log
r}+\frac{d\log T(r)}{d\log r}\right)r
\end{equation}
Hence, the mass depends on both the gas density and temperature
profiles. For isothermal gas, the observed surface brightness (which
can be accurately obtained from \rosat\ PSPC data), is directly related
to the gas density. The surface brightness outside the cooling flow
 usually follows a $\beta$-profile (Cavaliere \& Fusco-Femiano 1976) with a
fixed background $B$:
\begin{equation}
I(r)=I_0\left[1+\left(\frac{r}{a}\right)^2\right]^{-3\beta+\frac{1}{2}}
+ B
\end{equation}
Here $\beta=\mu m_p \sigma_r^2/kT_g$ is the ratio of energy per unit
mass in galaxies to the energy per unit mass in gas, and $\sigma_r$ is
the velocity dispersion.  The parameters $a$, $\beta$, and the
background $B$ are obtained from a least-squares fit to the X-ray
data, and the gas density profile is then given by
\begin{equation}
\label{eq:dens}
\rho_g(r)=\rho_0\left[1+\left(\frac{r}{a}\right)^2\right]^{-\frac{3}{2}\beta}
\end{equation}
The error introduced by assuming an isothermal gas in the density
profile calculation is not significant, since the fraction of the
bolometric luminosity emitted in the Snowden bands R5-R7 used in our
analysis (0.7-2.0 keV) varies little with temperature for all clusters
in our sample. 

The central density $\rho_0$ can be found from the surface brightness
profile as follows: from the known central surface brightness of the
cluster (the $\beta$ profile extrapolated to the cluster center), the
cluster redshift, gas temperature, abundance, absorbing hydrogen
column density, effective area of the \rosat\ mirrors and quantum
efficiency of the PSPC in the 0.7-2.0 keV range, we can calculate the
emission integral $EI=\int n_p n_e dV$. For an isothermal $\beta$
model the emission integral is:
\begin{equation}
EI=\pi^{3/2}\frac{n_e}{n_p}n_0^2 a^3\frac{\Gamma(3\beta-3/2)}{\Gamma(3\beta)}
\end{equation}
where $n_0$ is the central proton density, $\Gamma$ is the gamma
function, and for an assumed typical elemental abundance 0.3 Solar,
$n_e/n_p=1.17$ and $\rho=1.35m_p n_0$. The total gas mass can then be
found by integrating Equation~\ref{eq:dens} over the total volume. The
effect of the cooling flow on the total gas mass measured at radii of
1 Mpc or greater from the cluster core is $<10\%$, since most of the
gas mass resides in outer regions of clusters.

The most accurate temperature profiles for our sample clusters are
available for A426 (presented here), and A2199 (Markevitch et
al. 1999). In both cases, the temperature outside the cooling core can
be well approximated by a polytrope, $T\propto\rho_g^{\gamma-1}$ with
$\gamma \sim 1.2$.

For a gas distribution given by Equation~\ref{eq:dens}, the total mass enclosed
in a sphere of radius $r=x\,a$ is:

\begin{align}
\notag M(<r)&=\frac{k T(r)}{G m_p \mu}\frac{3\beta \gamma r^3}{a^2+r^2}=\\
& = 3.70\times 10^{13} M_{\odot} \frac{0.60}{\mu}\, \frac{T(r)}{\rm 1\;keV}\,\frac{a}{\rm 1\;Mpc}\, \frac{3 \beta \gamma x^3}{1+x^2}
\end{align}

Here $T$ is the real temperature, rather than a projection on the
plane of the sky. Markevitch et al. (1999) has shown that as long as
the temperature is proportional to a power of density, and the density
follows a $\beta$-model, the real temperature differs from the
projected temperature only by a constant factor, given by:
\begin{equation}
\frac{T_{\rm proj}}{T}=\frac{\Gamma\left[\frac{3}{2}\beta(1+\gamma)-\frac{1}{2}\right]\ \Gamma(3\beta)}{\Gamma\left[\frac{3}{2}\beta(1+\gamma)\right]\ \Gamma(3\beta-\frac{1}{2})}
\end{equation}
This correction factor is in the range 0.9 to 0.98 for all clusters in our sample.

\subsection{ROSAT data analysis}
Archival \rosat\ PSPC images were reduced using the standard analysis software (Snowden et al. 1994) that flat-fields the images and excludes periods of high particle background, as well as a period of 15 seconds after the high voltage is turned on. In order to maximize the signal-to-noise ratio, we use only Snowden energy bands R5-R7 corresponding to $\sim0.7-2.0$ keV.

We fit the surface brightness profiles with $\beta$ models, with the
core radius, $\beta$, the background, and the normalization as free
parameters. Since we are primarily interested in the gas properties at
large radii, the surface brightness profiles were fitted only outside
twice the cooling radii taken from White, Jones \& Forman (1997). An
acceptable $\chi^2$ cannot be obtained when the cooling flow region is
included. Point sources were excluded from all images manually.

The results of the fitting procedure are shown in
Table~\ref{tab:sbcomp}.  Our determinations agree very well with
earlier results from both \rosat\ (Vikhlinin et al. 1999) and
\einstein\ (Jones \& Forman 1999) observations.  

\begin{deluxetable}{lccccccccc}
\tablecaption{Surface brightness fitting results, comparison with
values from literature\label{tab:sbcomp}}
\tablewidth{7in}
\tablehead{Object  & \multicolumn{2}{c}{Our results} &
\multicolumn{2}{c}{Vikhlinin et al. 1999} & \multicolumn{2}{c}{Jones
\& Forman 1999} & \multicolumn{2}{c}{Markevitch et al. 1999}\\
\cline{2-3} \cline{4-5} \cline{6-7} \cline{8-9} & $\beta$ & $r_c$ &
$\beta$ & $r_c$ & $\beta$ & $r_c$ & $\beta$ & $r_c$}
\startdata
A262 & $0.53\pm0.03$ & $0.15\pm0.03$ & $\cdots$ & $\cdots$ & $0.55\pm0.05$ &
$0.09\pm0.03$ & $\cdots$ & $\cdots$ \nl
A426 & $0.58\pm0.02$ & $0.28\pm0.02$ & $\cdots$ & $\cdots$ & $0.55\pm0.03$ &
$0.28\pm0.05$ & $\cdots$ & $\cdots$ \nl
A478 & $0.75\pm0.02$ & $0.32\pm0.04$ & $0.76\pm0.11$ &
$0.30\pm0.13$ & $0.75\pm0.01$ & $0.31\pm0.03$ & $\cdots$ & $\cdots$  \nl
A1795 & $0.88\pm0.02$ & $0.41\pm0.03$ & $0.83\pm0.02$ &
$0.39\pm0.02$ & $0.73\pm0.08$ & $0.29\pm0.10$ & $\cdots$ & $\cdots$ \nl
A2052 & $0.65\pm0.03$ & $0.12\pm0.04$ & $0.64\pm0.02$ &
$0.10\pm0.05$ & $0.66\pm0.09$& $0.12\pm0.05$ & $\cdots$ & $\cdots$\nl
A2063 & $0.66\pm0.04$ & $0.20\pm0.03$ & $0.69\pm0.02$ &
$0.22\pm0.02$ & $0.62\pm0.05$ & $0.17\pm0.02$ & $\cdots$ & $\cdots$\nl
A2199 & $0.63\pm0.01$ & $0.12\pm0.01$ & $0.64\pm0.01$ &
$0.14\pm0.01$ & $0.62\pm0.05$ &$ 0.13\pm0.03$ & 0.636 & 0.134\nl
MKW4s & $0.64\pm0.10$ & $0.20\pm0.06$ & $\cdots$ & $\cdots$ &
$\cdots$ & $\cdots$ & $\cdots$ & $\cdots$ \nl
\enddata
\end{deluxetable}

\subsection{ASCA data analysis}

The \asca\ X-ray observatory (Tanaka, Inoue \& Holt 1994) spatially
resolved spectral data can be used to constrain the gas temperatures
at different regions of the clusters. The ASCA mirrors have an energy
and position dependent PSF that needs to be correctly taken into
account. Two independent methods that correct for the PSF (Churazov et
al. 1996; Markevitch et al. 1998) have been used in the past and were
found to be in very good agreement (Donnelly et al. 1998). The first
method approximates the \asca\ PSF as having a core and broad
wings. It uses the exact PSF correction for the core (inner 6`), and a
Monte Carlo correction for the scattered light in the wings of the
PSF.  The second method simultaneously fits temperatures in all
selected regions, taking into account the observed surface brightness
in each region and using the actual measured PSF.

\begin{figure*}[t]
\centerline{\includegraphics[height=4.0in]{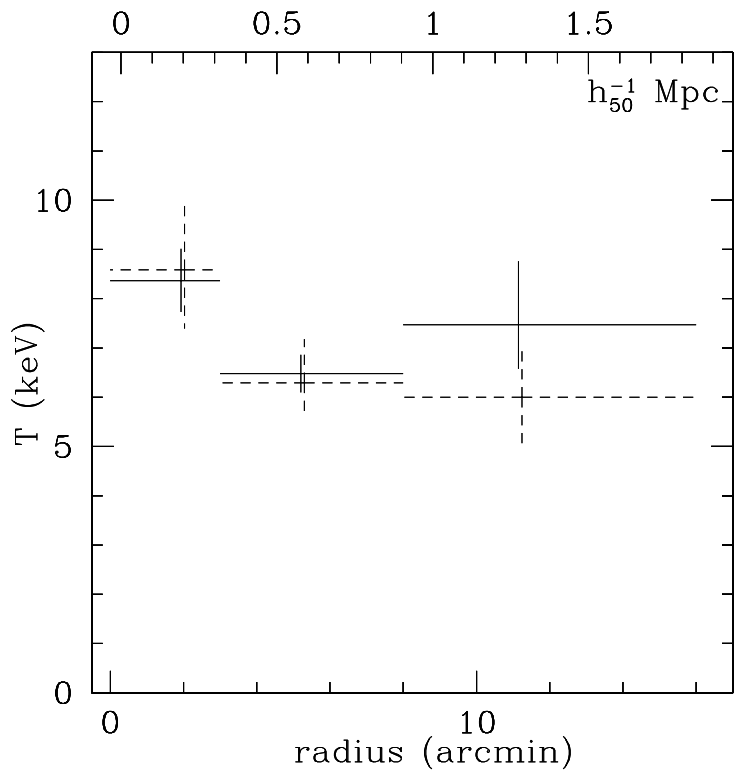}}
\caption{Our \asca\ temperature profile for A399 (solid line) shown for
comparison with temperature measurements from Markevitch et al. 1998
(dashed line). Our results are in good agreement. Note that the
slightly higher temperature we measured at a large cluster radius is
probably due to a small azimuthal asymmetry in the temperature
structure present in this cluster. The system may be interacting with
the nearby cluster A401.} \label{fig:399}
\end{figure*}

\begin{figure*}
\centerline{\rotatebox{-90}{\includegraphics[height=6.0in]{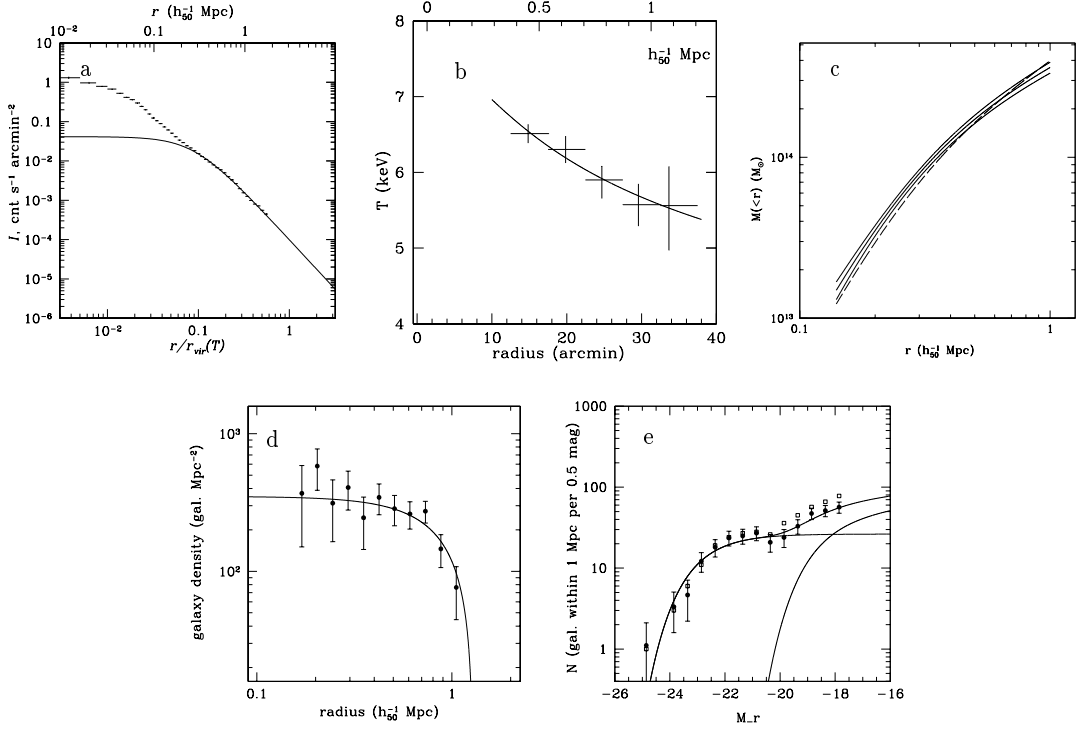}}}
\caption{Sample plots describing our analysis on A426. {\em a})
\rosat\ PSPC surface brightness profile fitted with a
$\beta$-profile. The region inside twice the cooling radius was
excluded from the fit.  {\em b}) \asca\ temperature profile for A426,
fitted with a polytropic function. The cooling flow region was
excluded.  {\em c}) Total mass profile for A426, with 90\% confidence
limits denoted by the thin lines. The dashed line is the corresponding
isothermal profile. {\em d}) Galaxy surface density profile of A426
fitted with a $\beta$ model with $\beta=2/3$. The core radius agrees
within the uncertainties with the X-ray core radius measured from the
PSPC surface brightness profile. {\em e}) Optical luminosity function
for A426, fitted with a sum of two
Schechter functions. \label{fig:426}}
\end{figure*}

We have adopted temperature profiles for three of the clusters (A478,
A1795 and A2199) previously generated by Markevitch et al. (1998,
1999) using the second method described above. For the remaining five
objects (A262, A426, A2052, A2063 and MKW4s), we have constructed
temperature profiles using the first method.  To check that the two
methods for generating temperature profiles are consistent, we have
constructed a temperature profile for A399 (Figure~\ref{fig:399}),
which was presented by Markevitch et al. (1998). Applying the two
methods yields results that agree well within their
uncertainties. (The slightly different temperatures measured at large
radii may be due to the small azimuthal asymmetry in the temperature
structure present in this cluster.)  A sample temperature profile
(A426) we generated using the first method and a corresponding total
mass profile obtained by fitting the temperature profile with a
polytropic function are shown in Figure~\ref{fig:426}.  A temperature
profile for A426 has also been measured by Eyles et al. (1991) using
an X-ray telescope flown on the \emph{Spacelab 2} mission and is in
excellent agreement with our measurement.

\begin{figure*}
\centerline{\rotatebox{-90}{\includegraphics[height=6.0in]{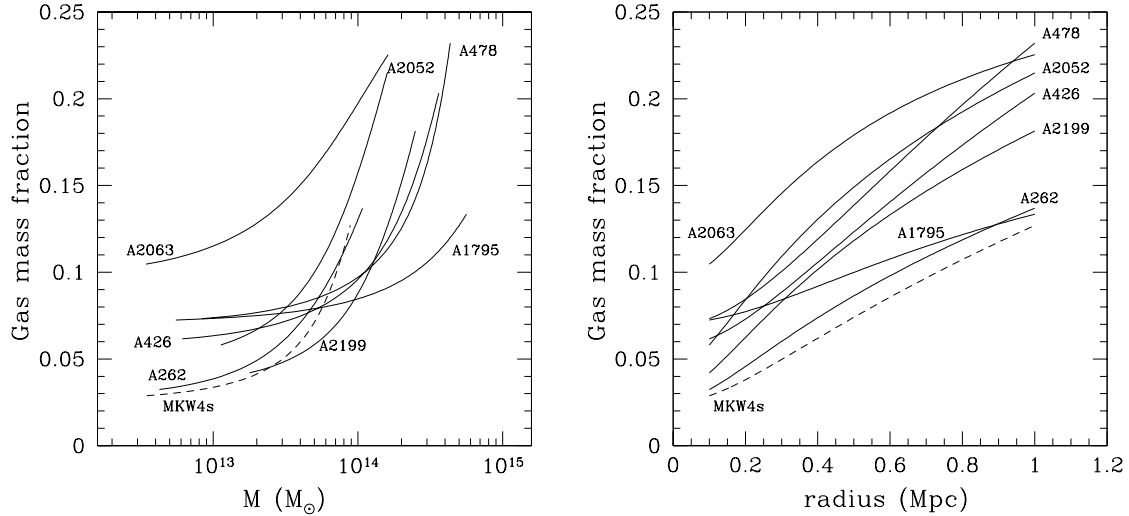}}}
\caption{Gas mass fractions ($f_{gas}$) ({\em left}) in the range
$0.1-1.0~$Mpc plotted as a function of enclosed mass, ({\em right})
plotted as a function of distance from cluster center in Mpc. Clusters
are shown as solid lines, the group (MKW4s) as a dashed line. Error
bars were omitted for clarity. The errors in $f_{gas}$ given in
Table~\ref{tab:tfits} for a radius of 1 Mpc are typical of the whole
range shown for each cluster.
\label{fig:gasmass}}
\end{figure*}

Gas mass fractions for all clusters in our sample are plotted as
functions of enclosed mass and radius in Figure~\ref{fig:gasmass} and
the results of our fitting are given in Table~\ref{tab:tfits}. The gas
mass fractions reach $\sim 0.15-0.25 \, h_{50}^{-3/2}$ at a radius of
1 Mpc.

\begin{deluxetable}{lccccc}
\tablecaption{Temperature and surface brightness fitting results -
Cluster properties\label{tab:tfits}}
\tablehead{Object & \multicolumn{1}{c}{$\rho_0$\tablenotemark{a}} & $\gamma$
&\multicolumn{1}{c}{$M$ (1~Mpc)} & \multicolumn{1}{c}{$M_{\rm gas}$
(1~Mpc)} & \multicolumn{1}{c}{$f_{\rm gas}$(1~Mpc)} \\ & $10^{13} M_{\odot}$ Mpc $^{-3}$ &
& $10^{13} M_{\odot}$ & $10^{13} M_{\odot}$ &}
\startdata
A262  & $3.97$ & $1.07\pm0.09$ & $10.8\pm1.4$ &
$1.47$ & $0.137\pm0.018$ \nl
A426  & $9.60$ & $1.14\pm0.06$ & $36.2\pm0.6$ & $7.35$ & $0.203\pm0.004$ \nl
A478  &  $15.7$ & $1.27\pm0.40$\tablenotemark{b} &
$43^{+16}_{-32}$ & $10.1$ & $0.23^{+0.08}_{-0.18}$ \nl
A1795 &  $10.0$ & $1.16^{+0.09}_{-0.12}$\tablenotemark{b} &
$56.3^{+11}_{-9.4}$ & $7.51$ & $0.133^{+0.027}_{-0.022}$ \nl
A2052 &  $21.6$ & $1.15\pm0.07$ & $15.9\pm1.7$ &
$3.42$ & $0.215\pm0.023$ \nl
A2063 &  $9.86$ & $\cdots$\tablenotemark{c} & $16.2\pm1.0$ &
$3.64$ & $0.225\pm0.014$\nl
A2199 & $24.1$ & $1.17\pm0.07$\tablenotemark{d}  &
$25.0\pm2.9$ & $4.53$ & $0.181\pm0.021$\nl
MKW4s & $27.2$ & $1.23\pm0.15$ & $8.9^{+2.0}_{-1.7}$ &
$1.13$ & $0.127^{+0.029}_{-0.024}$ \nl
\enddata
\tablenotetext{a}{Central gas density extrapolated from the best fit $\beta$-model.}
\tablenotetext{b}{Based on temperature profiles from Markevitch et al. 1998}
\tablenotetext{c}{Only one temperature is available outside the cooling flow
region due to the quality of the data. We assume an isothermal
temperature profile.}
\tablenotetext{d}{Based on temperature profile from Markevitch et al. 1999}
\end{deluxetable}


\section{Optical data reduction and analysis}
For measuring cluster luminosities, we use the Digitized Second
Palomar Sky Survey (DPOSS), calibrated with photometric CCD images
taken at the Palomar 60-in. telescope in the Gunn-Thuan \g, \r, and
\i\ bands (Weir et al. 1995a, Djorgovski et al. 1998).  

\subsection{Plate processing}
The conversion of photographic plate emulsion density to intensity
using the plate densitometry spots is described in Weir et
al. (1995b). 

The Sky Image Cataloging and Analysis Tool (SKICAT) has been developed
to detect objects and perform star/galaxy classification on both DPOSS
plates and CCD calibration data (Weir et al. 1995b). SKICAT is
presently optimized for measuring fainter objects than $m\simeq$ 16
mag. Clusters in our sample contain galaxies brighter than this limit;
hence we have used SExtractor (Bertin \& Arnouts, 1995) for detecting
objects and classifying stars and galaxies. 

\subsection{Photometric calibration}
\begin{figure*}
\centerline{\rotatebox{-90}{\includegraphics[height=6.0in]{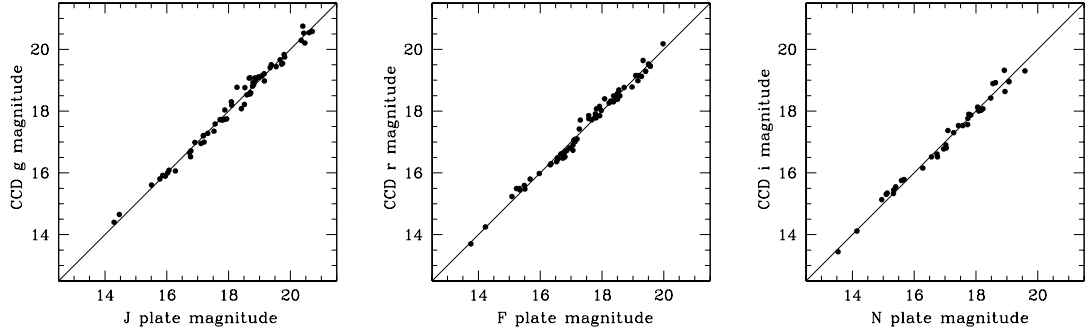}}}
\caption{Example of calibration transformation between plate and CCD
photometry for Abell 478. Only objects without overlapping isophotes
or close neighbors not resolved on the plate are used to derive a
calibration in order to avoid possible biases. The true scatter of the
calibration relation is slightly larger when \emph{all} matched
objects are included, and results from both problematic photometry in
crowded fields with numerous overlapping objects (edges of spiral
galaxies etc.), and the effect of slightly different bandpasses of the
CCD and plate data.\label{fig:transform}}
\end{figure*}

CCD images were obtained under photometric conditions for A262 (taken
on 13 Dec 1998), A426 (12 Feb 1995), A478 (18 Sep 1998), A1795 (18 Jul
1999), A2063 (12 Jul 1999), and A2199 (18 Jul 1999) in the Gunn-Thuan
\g, \r, \i\ bands. To provide photometric calibration for A2052 and
MKW4s, we use CCD images of different Abell clusters located on the
respective plates near the clusters of interest: A2063 to calibrate
A2052, A1495 (17 May 1998) to calibrate MKW4s. In order to correct the
calibration of A2052 and MKW4s for vignetting effects, we median
averaged $\sim100$ POSS-II fields to obtain a vignetting map. The
luminosity correction is on the order of a percent for both
clusters. An example of a calibration transformation derived for A478
is shown in Figure~\ref{fig:transform}.

In order to obtain the rest-frame galaxy luminosities, we need to 
correct for galactic absorption and k-dimming. Extinction corrections
for clusters in our sample are given in Table~\ref{tab:sample} , taken
from Schlegel et al. (1998).

K-corrections depend on spectral type, which can be related to galaxy
morphological type. Since an automated morphological classification of
galaxies in our sample is beyond the scope of this work, as well as
very problematic at faint magnitudes, we assume a morphological
composition and adopt k-corrections in a statistical manner
(Table~\ref{tab:kcor}). A sample of 55 nearby rich clusters in the
redshift range of our interest has been studied (Dressler 1980;
Whitmore, Gilmore \& Jones 1993, Dressler et al. 1997) and the
morphological fractions determined as a function of the density of the
environment. We follow Dressler et al. (1997) and adopt the following
morphological fractions to be typical of the clusters in our sample:
25\%:40\%:35\% for E:S0:Sp. The k-corrections are calculated using
model galaxy SEDs from Small (1996) for different galaxy morphological
types and the Gunn-Thuan \g, \r, \i\ filter bandpasses (Weir et
al. 1995a). We found the k-corrections from Small (1996) to be in
agreement with an independent study by Fukugita et al. (1995).

\begin{deluxetable}{ccccc}
\tablecaption{K-corrections - Statistically combined\label{tab:kcor}}
\tablehead{Object & \multicolumn{3}{c}{K-correction} \\ 
  & Gunn \g & Gunn \r & Gunn \i}
\startdata
A262  & 0.02 & 0.02 & 0.01 \nl
A426  & 0.02 & 0.02 & 0.01 \nl
A478  & 0.14 & 0.07 & 0.06 \nl
A576  & 0.05 & 0.03 & 0.03 \nl
A1795 & 0.08 & 0.04 & 0.04 \nl
A2052 & 0.05 & 0.02 & 0.02 \nl
A2063 & 0.05 & 0.03 & 0.03 \nl
A2199 & 0.04 & 0.03 & 0.02 \nl
MKW4s & 0.04 & 0.03 & 0.02 \nl
\enddata
\end{deluxetable}


It should be noted that for the majority of clusters in our sample,
the k-correction in all three bands (\g, \r, and \i)
and all morphological types is no larger than 0.14 mag, and for about
half of our sample, the k-correction is below 0.05 mag. Therefore, the exact morphological fraction is not critical, particularly in the \r\ and \i\ bands, where the differences in the
k-corrections between the different morphological types are quite
small at low redshifts. For example, had we assumed a spiral fraction
of 20\% rather than 35\%, the statistically combined k-correction
would change by no more than 0.02 mag for all clusters in our sample.
Evolutionary effects also are insignificant due to the low redshift of
our sample.

To convert apparent magnitudes to absolute magnitudes, we use the standard
relation:
$$
M=m-DM-E-k
$$
where $DM$ is distance modulus, $E$ is the galactic absorption and $k$ is the
k-correction. Assuming $q_0=0.5$, the distance modulus, $DM$, is:
$$
DM=43.89+5\log(z)-5\log(h_{50})+0.54 z
$$

A deceleration parameter $q_0=0$ would increase the distance modulus
$DM$ by $0.03$ mag at $z=0.05$ and by $0.08$ mag at $z=0.15$. A mean
cluster redshift error of $0.5\%$ results in an absolute magnitude error of
$\sim0.01$, well below other random and systematic errors. 

For comparison with other studies, we convert our Gunn \g\ magnitudes to the Johnson $V$ band using the relation $V=g-0.03-0.42 \ (g-r)$ (Windhorst et al. 1991). A mean $g-r$ color for low redshift clusters is $g-r=0.3$, giving $V=r+0.14$

\subsection{Luminosity function determination}
The values of galaxy cluster and group luminosities (\eg\ Oemler 1974, Dressler 1978a,b; Bucknell et al. 1979; Lugger 1986; Oegerle et al. 1986; Ferguson \&
Sandage 1990) used in previous mass-to-light ratio studies
 date back to the first generation Palomar Sky Survey
plates, or plates of similar grade taken elsewhere. In many
studies only a small number of objects was used for photometric
calibration, and star-galaxy classification was performed using simple
two-parameter classifiers that are outperformed by more recent methods. In some studies object detection was performed by visual inspection. 

Cluster luminosity functions have recently been studied using photometric CCD images (\eg\ Lopez-Cruz et al. 1997). However, the number of clusters thus studied is still small, and at low redshifts the volume sampled is limited. Photographic plates still remain the optimal way of studying large samples of clusters over large areas of the sky.

Here we present the luminosity functions (LFs) for our sample of 8 relaxed clusters and one group, obtained from the digitized second generation
Palomar Sky Survey plates, calibrated with CCD images in the
Gunn-Thuan \g, \r, \i\ system, and sampling 1 Mpc from the cluster centers.  

\subsubsection{Background subtraction}
Different LF studies have taken different paths in estimating
background counts. Some have used values obtained in other independent
studies, where different filters, angular coverage, or a different
definition for galaxy magnitudes were used. These factors introduce
errors that were estimated to contribute to a total uncertainty of
$\pm50\%$ in the background correction (Oemler 1974, Lugger 1986,
Colless 1989). Lower values in the background uncertainty were reported by Dressler (1978a) using Shane and Wirtanen counts (25\% variation). Lopez-Cruz (1995) found a similar variation in $R$-band counts on scales $\sim0.4^{\circ}$ 

We have analyzed the background galaxy counts on POSS-II plates 725, and found a 19\% variation on scales of $0.5^{\circ}$ with a limiting \emph{J} magnitude $19.5$. This is in agreement with the findings of Dressler (1978a) and the expected variation in the angular covariance function (Groth \& Peebles 1975) on scales of 0.5\deg. We thus assume the error in the background counts $N$ is the maximum of $\sqrt{N}$ and $N/5$.

Assuming Poisson uncertainties in the uncorrected galaxy counts, the error in the corrected counts is given by 
$$
\sqrt{N+\rm{max}\left(\sqrt{\emph{N}};\emph{N}/5\right)^2}.
$$

The background subtracted differential LFs were fitted with the commonly used
Schechter function (Schechter 1976):
$$
n(L)dL=N^*(L/L^*)^{\alpha}\exp(-L/L^*)d(L/L^*)
$$

Once the parameters $L^*$, $N^*$ and $\alpha$ have been determined, the total
cluster luminosity is given by:
$$
L_{\rm clus}=\int_{0}^{\infty}Ln(L)dL=N^*\Gamma(\alpha+2)L^*.
$$
For the three lowest redshift clusters (A262, A426, A2199) where the absolute magnitude range sampled is the greatest, we found a sum of two Schechter functions greatly improves the chi-square of the fit. In such cases the slope of the brighter component was fixed at $\alpha=-1$, the remaining parameters were left free.

Since we measure cluster masses within 1~Mpc from the cluster centers, to obtain the corresponding luminosities over these cluster volumes, we must correct for outlying cluster galaxies projected near the cluster center. To calculate this correction we need to know the galaxy number density and average galaxy luminosity as a function of distance from the cluster center. We fitted the galaxy number density profiles with $\beta$ models and found the coefficients to have a larger uncertainty, but to be consistent with the gas density fitting results. We thus assume that the distribution of galaxies follows the distribution of intracluster gas, and that the average galaxy luminosity is independent of the density of the environment. The correction factor is in the range $0.90-0.97$ and has the effect of decreasing the true total cluster luminosities. Some studies have suggested both of the underlying assumptions may be violated. However, the error this may introduce can be only of the order of a percent, since we sample to a radial distance about 5 times the typical cluster core radius.

In our determinations of cluster luminosities we assume the Schechter function is a universal LF valid over a large magnitude range from giant galaxies to dwarfs. We study the LFs of galaxies with apparent \r\ magnitude brighter than $\sim19$ mag (corresponding to an absolute magnitude $M_r \sim-17.5$ to $-19.5$ depending on the cluster redshift). The giant galaxies contribute most of the cluster luminosity. Typically, galaxies with $M_r<-19$ comprise 80-90\% of the total cluster luminosity, with the exact number depending on $M^*$ and $\alpha$. 

Some studies have found the Schechter function does not describe the
cluster LF well at the faint end. Trentham (1998) studied \B-band LFs
of 9 Abell clusters and showed that LFs tend to flatten for
$-18<M_B<-16$ and then rise for fainter galaxies, with slopes varying
in the range $-1.3<\alpha<-1.8$. However, for our purposes this effect
is negligible, since the dwarf galaxies contribute only a small
fraction of the total light. Assuming the LF is described by a
Schechter function with $M^*\sim-22$ and $-1.4<\alpha<-1.0$ for
$M_r<-17$ (as suggested by Trentham 1998), the effect of a faint end
slope varying in the range $-1>\alpha>-2$ results in a negligible
change in the total cluster luminosity ($\sim1\%$).  The results of
the fitting procedure are shown in Table~\ref{tab:lf}.

\begin{deluxetable}{lll}
\tablecaption{\sc Luminosity function fitting results\label{tab:lf}}
\tablehead{Object & $L_V$ (1Mpc) $L_{\odot}$ & $M/L_V$ (1Mpc)}
\startdata
A262 & 12$\pm4\times10^{11}$ & $90\pm32$ \nl
A426 & $47\pm21\times10^{11}$ &  $77\pm$34\nl
A478 & $31\pm8\times10^{11}$ & $138^{+62}_{-109}$ \nl
A1795 & $24\pm6\times10^{11}$ & $234^{+74}_{-70}$\nl 
A2052 & $16\pm4\times10^{11}$ & $99\pm27$ \nl
A2063 & $17\pm4\times10^{11}$ & $95\pm23$ \nl
A2199 & 17 $\pm6\times10^{11}$ & $147\pm55$ \nl
MKW4s & $8.8\pm3.2\times10^{11}$ & $101\pm43$  \nl
\enddata
\tablenotetext{}{Values are for $H_0=50~{\rm km}~{\rm s}^{-1}~{\rm
Mpc}^{-1}$.} 
\end{deluxetable}


\section{Mass-to-light ratios and constraints on $\Omega$}

The mass to light ratio, ${\rm M}/{\rm L}$, is used to parameterize
the amount of dark matter on various scales. ${\rm M}/{\rm L}$
increases with scale from galaxies to groups and clusters (Bahcall,
Lubin, \& Dorman 1995). However, a flattening of the ${\rm M}/{\rm L}$
vs. scale relation has been observed on scales beyond clusters, as
discussed in the introduction. Assuming that the mass-to-light ratios
of clusters are representative of the whole Universe, the mass density
of the Universe can be calculated from the observed mean luminosity
density of the Universe and ${\rm M}/{\rm L}$ of clusters.

The median ${\rm M}/{\rm L}$ of our sample is ${\rm M}/{\rm L_V}\sim
100 \, h_{50}~{\rm M_{\odot}}/{\rm L_{\odot}}$ (Table~\ref{tab:lf}). The mean
luminosity density of the universe is $\sim1\times 10^8 \, h_{50}~{\rm
L}_{\odot}{\rm Mpc}^{-3}$ (Efstathiou et al. 1988). This gives a
universal mass density of $\rho_m\simeq7\times10^{-31} \,
h_{50}^2~{\rm g}~{\rm cm}^{-3}$. With a critical density of $\rho_{\rm
crit}\simeq5\times10^{-30} \, h_{50}^2~{\rm g}~{\rm cm}^{-3}$, we
obtain $\Omega_0\simeq0.15$.

Our mass-to-light ratios within 1 Mpc are slightly lower than previous
results that used X-ray mass estimates: ${\rm M}/{\rm{L}_V}\sim 100 \,
h_{50}$ solar units compared to $\sim 120-150 \, h_{50}$ (Cowie 1987,
David et al. 1995). This discrepancy may be due in part to the
inability in previous work to correct for temperature structure. With
the advent of spatially resolved spectral measurements from \asca\, we
would expect a difference in the results especially if temperature
gradients are common. Second, we have used larger datasets for
calibrating the plate magnitudes, in comparison with earlier
studies. As a result, we expect the combined photometric properties of
larger samples of galaxies (such as the total luminosity) to be more
accurate estimates of the true values. There is only one cluster
(A262) studied both in this paper, and by David et al. (1995). The
mass-to-light ratios measured are in good agreement. 

${\rm M}/{\rm L_V}$ for most clusters in our sample are also lower
than estimates based on the virial mass estimator, which typically
yield ${\rm M}/{\rm L_V}\sim 125-180 \, h_{50}~{\rm M_{\odot}}/{\rm
L_{\odot}}$ (\eg\ Girardi et al. 1999, Carlberg et al. 1996). As
argued in the introduction, virial mass estimates can be misleading if
substructure is present, the assumption that mass follows light fails,
or when the volume sampled does not extend to the virial radius, which
is the case in many studies.

\begin{figure*}
\centerline{\includegraphics[height=4.0in]{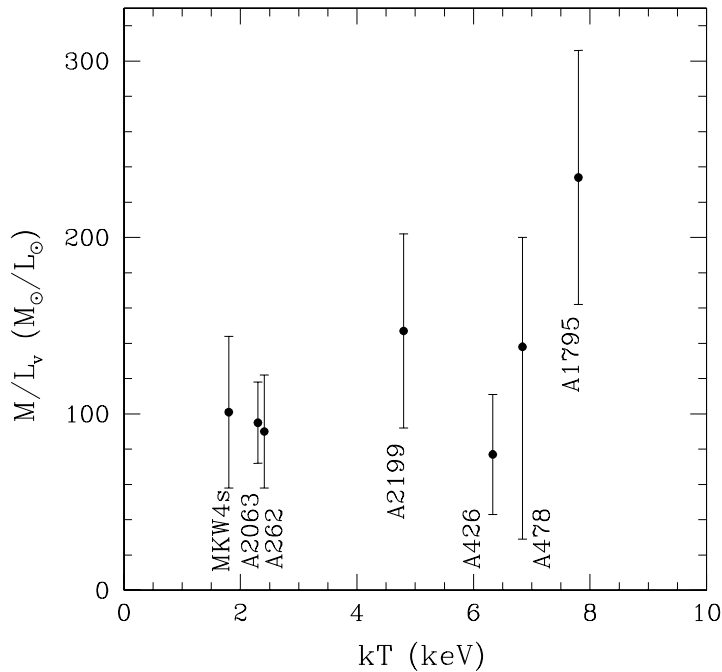}}
\caption{Mass-to-light ratios within 1 Mpc. The error bars include
uncertainties in both the mass and luminosity
determinations. $H_0=50~{\rm km}~{\rm s}^{-1}{\rm Mpc}^{-1}$ is
assumed.\label{fig:mlratio}} 
\end{figure*}


Our analysis shows that ${\rm M}/{\rm L}$ is roughly independent of
cluster mass as characterized by richness or temperature
(Figure~\ref{fig:mlratio} and Table~\ref{tab:sample}.) This is
contrary to the popular belief that mass-to-light ratios increase with
richness from groups to clusters, and is in agreement with the findings of David et al. (1995).

Standard Big Bang nucleosynthesis limits the baryon density of the
universe to $\Omega_b=0.076\pm0.004 \, h_{50}^{-2}$, where
$\Omega_b=f_b\Omega_0$, $f_b$ is the baryon mass fraction (Walker et
al. 1991, White et al. 1993, Tytler et al. 1996, Kirkman et al. 2000).
In Section 2 we showed the gas mass fraction reaches $\sim0.15-0.25 \,
h_{50}^{-3/2}$ at a 1 Mpc radius and tends to increase further towards
larger radii, with stars contributing only a few percent of the baryon
mass throughout. If we assume the standard Big Bang nucleosynthesis
calculations correctly predict the expected baryon fraction and that
the gas fraction found in clusters of galaxies is representative of
the baryon fraction in the Universe, as White et al. (1993) and David
et al. (1995) have done, we also can place an upper limit on
$\Omega_0$: $\Omega_0<0.076~f_b^{-1}~h_{50}^{-1/2}$. For our best
estimate $f_b=0.25$ (taking the upper limit to account for gas
fractions increasing beyond the region surveyed), we obtain
$\Omega_0<0.30~h_{50}^{-1/2}$, which is consistent with the constraint
on $\Omega_0$ from mass-to-light ratios for a presently favored value
of $H_0=65~{\rm km}~{\rm s}^{-1}{\rm Mpc}^{-1}$. We note that a larger
$\Omega_0$ is allowed if a lower gas mass fraction is adopted. For our
lower limit $f_b=0.15$, we obtain $\Omega_0<0.51~h_{50}^{-1/2}$.

\section{Conclusion}

We have investigated several fundamental properties of a sample of 7 Abell clusters and one group. We have utilized the Digitized Second Palomar Sky Survey optical data and photometric CCD images for constraining cluster luminosities, along with \rosat\ X-ray data and \asca\ spectra for constraining total and gas masses.

We have measured the median cluster mass-to-light ratios within 1 Mpc to be ${\rm M}/{\rm L_V}\sim 100 \, h_{50}~{\rm M_{\odot}}/{\rm L_{\odot}}$, corresponding to $\Omega_0\simeq0.15$. This is slightly lower than found in other studies that used X-ray mass estimates, and lower compared to results based on virial mass estimates. 

We have measured the gas mass fractions in the range 0.1-1 Mpc, and found
these to approach $0.15-0.25 \, h_{50}^{-3/2}$ towards the cluster
virial radii. Using the standard Big Bang nucleosynthesis
calculations, assuming that the baryon fraction seen in clusters to
within the virial radius of clusters is representative of the overall
baryon fraction in the Universe, we find the total matter density of the universe to be $\Omega_0<0.30 h_{50}^{-1/2}$.

Our two determinations of $\Omega_0$ are in agreement. Our results also are consistent within their uncertainties with other independent measurements of $\Omega_0$, such as the evolution of cluster abundance as a function of redshift (Bahcall 1999), microwave background fluctuations based on the COBE satellite results assuming both OCDM and LCDM models (\eg\ Cay\'{o}n et al. 1996), or measurements using distant supernovae (\eg\ Perlmutter et al. 1999, Riess et al. 1998)

As we enter the era of large format optical CCDs, it will become
possible to study the luminosity functions of large samples of
clusters out to the virial radii and reaching fainter magnitudes than 
photographic plates. New X-ray missions with better spectral and
spatial resolution, such as Chandra and XMM, will better constrain the
properties of the ICM in clusters, which will decrease the present
uncertainties on the mass-to-light ratios and limits on the baryon fraction.

\acknowledgments

V.~H. was partially supported by the Caltech SURF fellowship. V.~H. would like to thank the Harvard-Smithsonian center for Astrophysics for hospitality. C.~J. and R.H.D. acknowledge support form the Smithsonian Institute and NASA contract NAS8-39073. The DPOSS cataloging effort is supported by a generous grant from the Norris foundation. V.~H. would further like to thank A. Vikhlinin and M. Markevitch for useful comments.


\end{document}